\documentclass[12pt]{iopart}

\usepackage{epsfig}

\begin{document}

\title{RHIC Physics with the Parton Cascade Model}
\author{Steffen A.~Bass\dag\ddag, Berndt M\"uller\dag\ and 
	Dinesh K. Srivastava\S}
\address{\dag\ Department of Physics, Duke University, 
             Durham, North Carolina 27708-0305, USA}
\address{\ddag\ RIKEN BNL Research Center, Brookhaven National Laboratory, 
             Upton, New York 11973, USA}
\address{\S\ Variable Energy Cyclotron 
Centre, 1/AF Bidhan Nagar, Kolkata 700 064, India}            

\ead{bass@phy.duke.edu}

\submitto{\JPG}
\pacs{25.75.-q,12.38.Mh}

\begin{abstract}
We present an analysis of the net-baryon number rapidity distribution
and of direct photon emission in the framework of the Parton Cascade Model.
\end{abstract}


Collisions of heavy nuclei at relativistic energies are expected to 
lead to the formation of a deconfined phase of strongly interacting 
nuclear matter, often referred to as a Quark-Gluon-Plasma (QGP). 
Evidences for several of the signatures for the formation of this novel 
state of matter have recently been reported by experiments conducted 
at the Relativistic Heavy Ion Collider (RHIC) at Brookhaven National 
Laboratory \cite{rhic_highlights}. Many aspects of the experimental
data indicate that an equilibrated state of hot and dense matter is
formed in the collisions of Au nuclei at RHIC. However, it has not
yet been well established
how quickly this thermalized state is formed and which 
mechanisms are responsible for the rapid equilibration. 
It is thus of particular interest to identify processes that can give
information about the pre-equilibrium dynamics in these collisions.
In our contribution we focus on two such probes of the 
early phase at RHIC: direct photons and the
distribution of net baryon number. 

The parton cascade model~\cite{GM92} (PCM) provides a suitable framework 
for the study of the formation of a hot and dense partonic phase, 
starting from clouds of valence quarks, sea quarks, and gluons which 
populate the nuclei. The PCM was devised as a description of the early, 
pre-equilibrium phase of a nucleus-nucleus collision at high energy. 
The current implementation~\cite{VNIBMS} does not include a description 
of the hadronization of the partonic matter and of the subsequent 
scattering among hadrons. These late-stage processes, however, are not 
expected to significantly alter the distribution of net baryon number 
with respect to rapidity, since the net baryon number is locally 
conserved and baryon diffusion in a hadronic gas can be shown to be 
slow \cite{Stephanov}.   

The PCM assumes that the state of the dense partonic system can be 
characterized by a set of one-body distribution functions 
$F_i(x^\mu,p^\alpha)$, where $i$ denotes the flavor index 
($i = g,u,\bar{u},d,\bar{d},\ldots$) and $x^\mu, p^\alpha$ are 
coordinates in the eight-dimensional phase space. The partons are 
assumed to be on their mass shell, except before their first interaction. 
In our numerical implementation, the {\sc GRV-HO} parametrization
\cite{grv} is used, and the parton distribution functions are sampled 
at an initialization scale $Q_0^2$ to create a discrete set of particles. 
Partons generally propagate on-shell and along straight-line trajectories 
between interactions. Before their first collision, all partons move with 
the beam (target) rapidity and do not have an ``intrinsic'' transverse 
momentum.

The time-evolution of the parton distribution is governed by a 
relativistic Boltzmann equation:
\begin{equation}
p^\mu \frac{\partial}{\partial x^\mu} F_i(x,\vec p) = {\cal C}_i[F]
\label{eq03}
\end{equation}
where the collision term ${\cal C}_i$ is a nonlinear functional 
of the phase-space distribution function. The calculations discussed
below include all lowest-order QCD scattering processes between 
massless quarks and gluons. A low momentum transfer cut-off 
$p_T^{{\rm min}}$ is needed to regularize the infrared divergence 
of the perturbative parton-parton cross sections.  Additionally, we 
include the branchings $q \to q g$, $q \to q\gamma$, $ g \to gg$ and 
$g \to q\overline{q}$ \cite{frag}.  The soft and collinear singularities 
in the showers are avoided by terminating the branchings when the 
virtuality of the time-like partons drops below $\mu_0 = 1$ GeV. 
The results to be discussed below have been obtained using the 
{\sc VNI/BMS}~\cite{VNIBMS} implementation of the PCM.

\begin{figure}[tb]   
\centerline{\epsfig{file=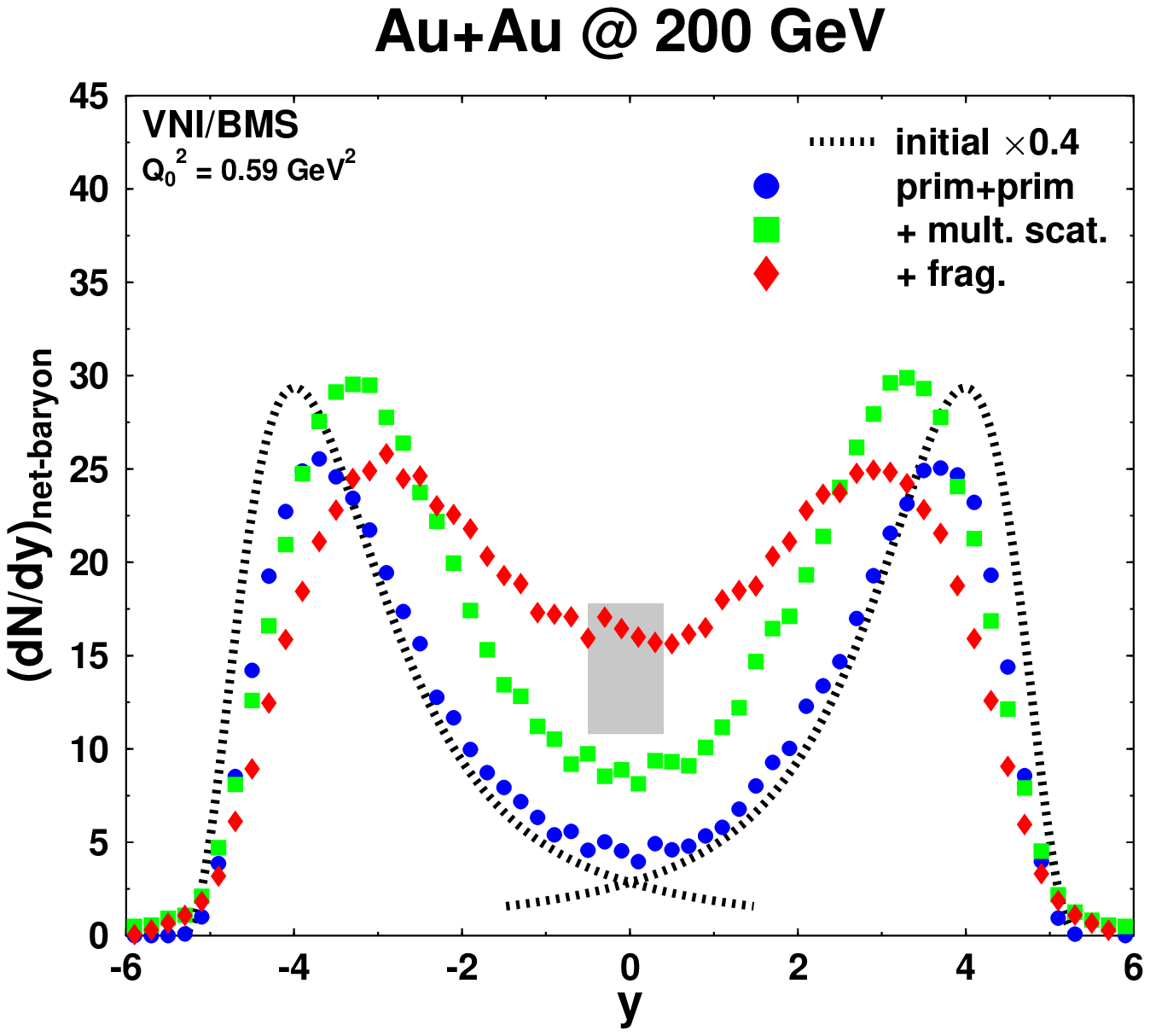,width=8cm}
\epsfig{file=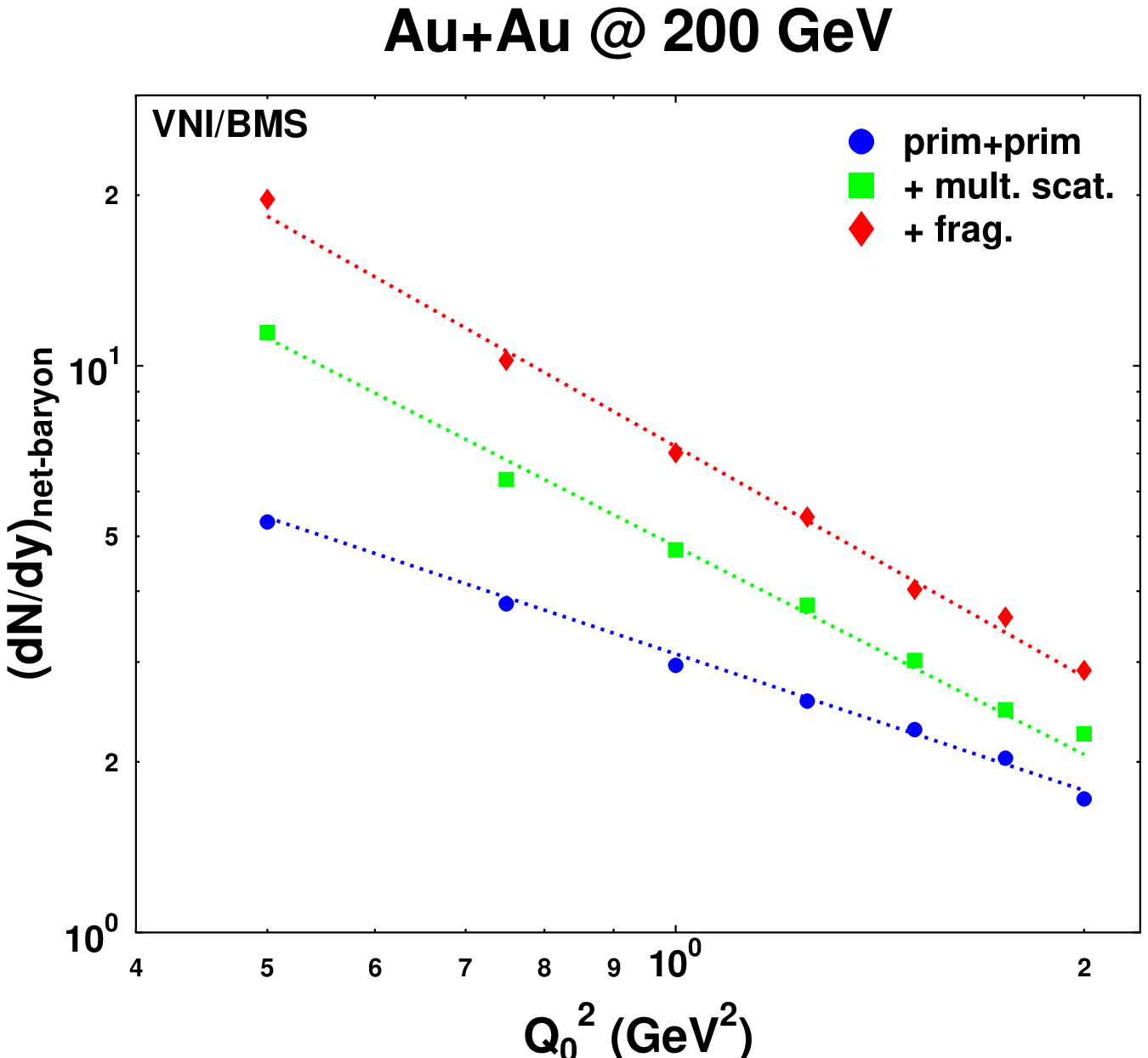,width=8cm}}
\caption{Net-baryon rapidity distribution (left) and initialization scale 
and cut-off 
dependence of the net baryon density at mid-rapidity in the PCM (right).}
\label{fig1}
\end{figure}

The left panel of Figure~\ref{fig1}  
shows the PCM prediction for the net baryon
rapidity distributions for $\sqrt{s}_{NN}=200$~GeV.
Solid circles in Fig.~\ref{fig1} denote a calculation in which the
PCM has been restricted to primary-primary parton scatterings,
and therefore reflects a calculation in which each parton is
allowed to scatter only once. Already one hard collision is
sufficient to deposit a net surplus of quarks into the mid-rapidity
region, resulting in a net baryon density at $y_{cm}=0$ 5.0 at 200~GeV. 
For comparison, the net baryon
number distribution for each colliding nucleus, scaled by a factor
0.4 from the distribution shown in Fig.~\ref{fig1}, is shown as
a dotted line. This initial state rapidity distribution can
be calculated from the parton structure functions by relating
the longitudinal momentum of the partons $p_z$ to the rapidity 
variable $y = Y + \ln x + \ln(M/Q_s)$,
where $Y$ is the rapidity of the fast moving nucleon, $M$ is the
nucleon mass, and $Q_s$ denotes the typical transverse momentum
scale. 

The remarkable similarity between the scaled initial
state rapidity distribution and the PCM calculation involving 
primary-primary scattering demonstrates that the
net baryon number distribution produced by first parton-parton
collisions is predetermined by the initial parton structure of
the nuclei. The factor 0.4 is the average ``liberation factor'' $c$
for partons in the PCM for the selected parameters.

The squares represent a calculation with
full parton-parton rescattering. Allowing for multiple parton
collisions increases the net baryon density at mid-rapidity roughly
by 75\%, filling up the
dip around mid-rapidity. This trend continues when parton fragmentation
is included (diamonds):  the net baryon density increases to
nearly 17. The rapidity change of a quark in each subsequent collision
after its liberation in the first hard scattering yields an average
rapidity shift of roughly 0.65 units per collision.
The band around mid-rapidity denotes the range of
experimental estimates for the net-baryon density at 
mid-rapidity \cite{bbar_exp,brahms}.

The right frame of Figure~\ref{fig1} 
shows the net baryon density at mid-rapidity as a
function of the momentum cut-off $p_T^{\rm{min}}$. The
observed power law dependence of the net baryon density as a function
of $Q_0$ stems from the properties of the pQCD cross sections in the
PCM. The absence of a saturation at small values of $Q_0$ indicates
that not all valence quarks are ``liberated'' in the range of cut-off values
considered here. Indeed, we find that the liberation factor for quarks
in the nuclear parton distributions varies from about 0.7 for $x>0.1$
to about 0.2 for $x\approx 0.01$.

Overall we find that
the valence quark distribution in the 
nucleon, combined with these multiple scattering effects, can explain
the net baryon excess observed in Au+Au collisions at RHIC in the
central rapidity region \cite{netb1}.

\begin{figure}[tb]   
\centerline{\epsfig{file=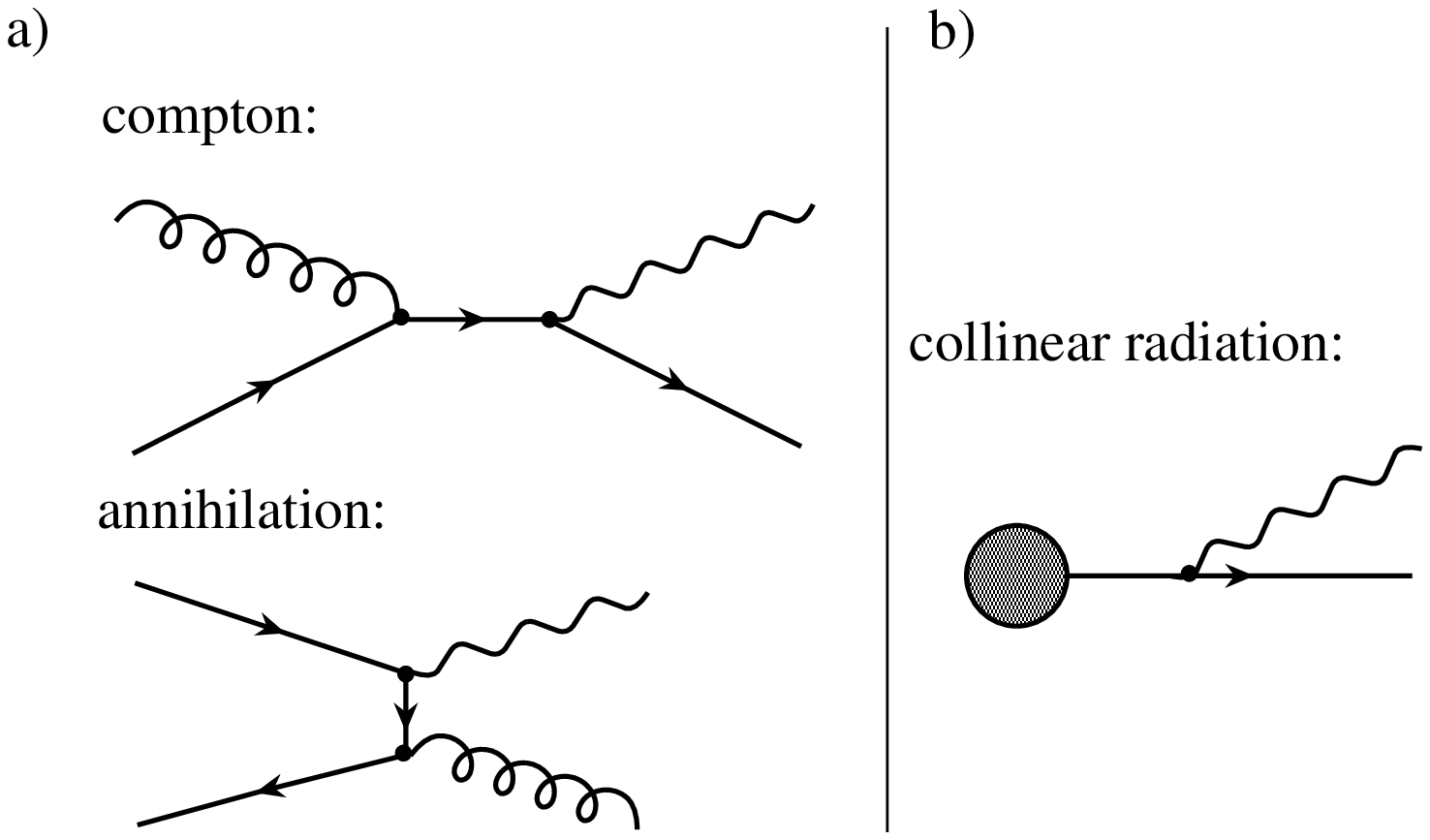,width=10cm}
\epsfig{file=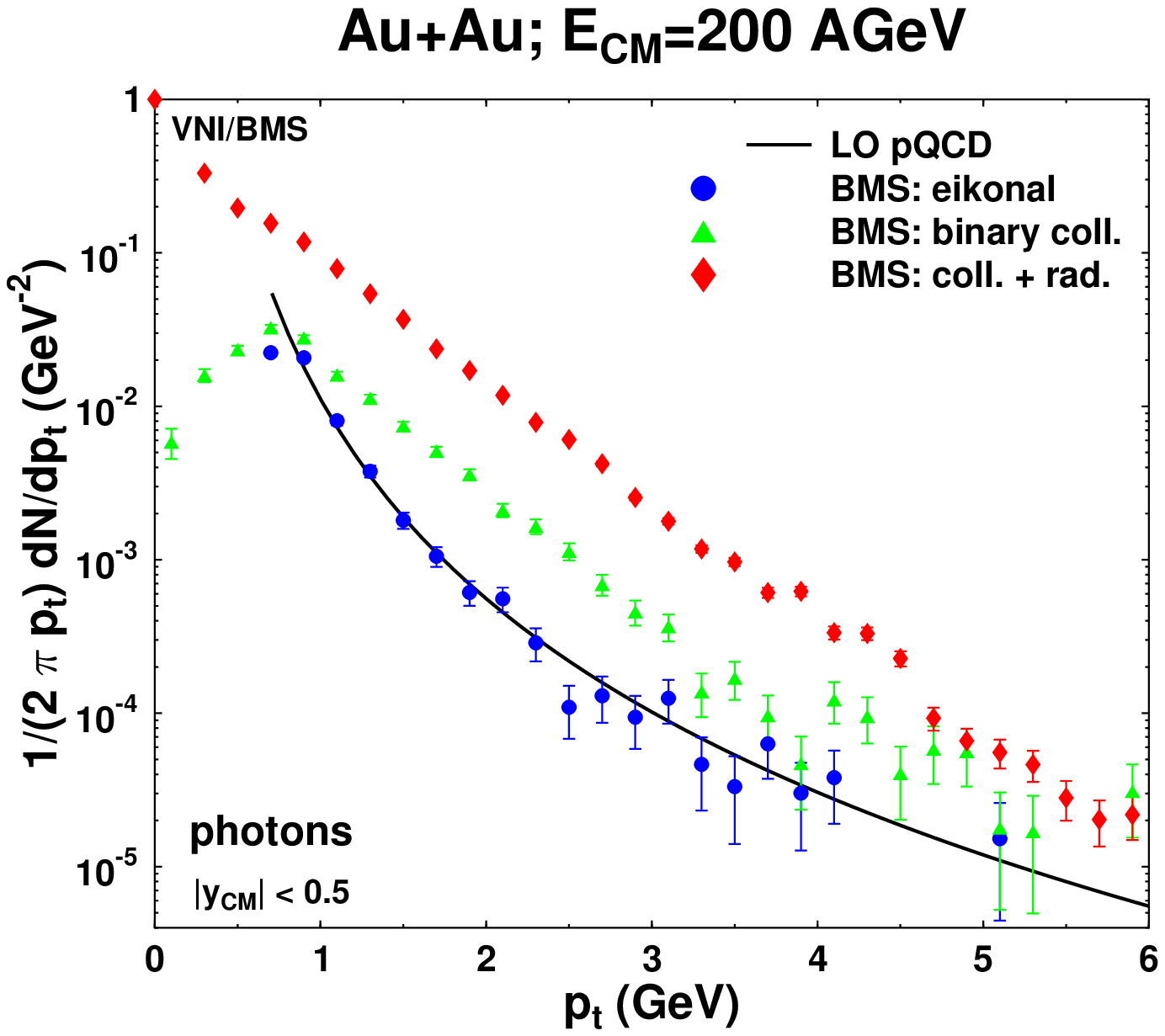,width=7cm}}
\caption{Left: photon productions processes included in the PCM.
Right: transverse momentum spectra of direct photons calculated in the PCM.}
\label{fig2}
\end{figure}

Figure~\ref{fig2} shows the photon production processes included in the 
PCM (left) as well as the
transverse momentum distribution (right) of direct photons 
calculated in the PCM.
The solid curve
gives a pQCD prediction for prompt photon production, due to
Compton and annihilation processes. The accuracy of the PCM
implementation is confirmed by the quantitative agreement between
this calculation and our results (shown as circles), when an eikonal
approximation is used to treat all the collisions in the cascade.
We note that a substantial enhancement 
of the high energy photons is caused by
the multiple scattering of partons (triangles).
Fragmentation processes lead to an even stronger enhancement in the production
of photons. Firstly, there is production of photons from the
fragmentation off the scattered quarks. Secondly, and more importantly,
the fragmentation leads to a rapid multiplication of partons and
enhanced multiple scatterings. The resulting calculation is denoted
by the diamonds in figure~\ref{fig2}. 
In summary, we find that 
multiple scattering of partons 
leads to a substantial production of high energy photons,
which rises further when parton multiplication due to
final state radiation is included \cite{bms_photon}. 
The photon yield is found to scale as $N_{\rm{part}}^{4/3}$ and
is directly proportional to the number of hard parton-parton collisions
in the system, providing
valuable information on the pre-equilibrium reaction dynamics of the
system.

\ack  
This work was supported in part by RIKEN, the Brookhaven National 
Laboratory, and DOE grants DE-FG02-96ER40945 and DE-AC02-98CH10886. 
S.A.B. acknowledges support from a DOE Outstanding Junior Investigator
Award.

\end{document}